\documentclass[runningheads]{llncs}
\usepackage[T1]{fontenc}
\usepackage{graphicx}
\usepackage{todonotes}
\usepackage{hyperref}
\usepackage{xcolor}

\usepackage{draftwatermark}

\usepackage{color}

\begin{document}
\SetWatermarkText{Pre-print}
\title{JELAI: Integrating AI and Learning Analytics in Jupyter Notebooks}
\titlerunning{Introducing JELAI}
%
\author{Manuel Valle Torre\inst{1}\orcidID{0000-0002-0456-8360} \and
    Thom van der Velden\inst{1} \and \\ 
    Marcus Specht\inst{1}\orcidID{0000-0002-6086-8480} \and
    Catharine Oertel\inst{1}\orcidID{0000-0002-8273-0132} 
}
\authorrunning{Valle Torre et al.}
%
\institute{Delft University of Technology, Delft, Netherlands \and
    \email{\{m.valletorre,thomvanderveld,m.m.specht,c.r.m.m.oertel\}@tudelft.nl}\\
    \url{https://www.tudelft.nl/en/eemcs/}
}
\maketitle              
\vspace{-0.2cm}
\textcolor{red}{\tiny The final version of this paper will be published in AIED 2025}
\vspace{-0.2cm}
\begin{abstract}
Generative AI offers potential for educational support, but often lacks pedagogical grounding and awareness of the student's learning context. 
Furthermore, researching student interactions with these tools within authentic learning environments remains challenging. 
To address this, we present JELAI, an open-source platform architecture designed to integrate fine-grained Learning Analytics (LA) with Large Language Model (LLM)-based tutoring directly within a Jupyter Notebook environment. 
JELAI employs a modular, containerized design featuring JupyterLab extensions for telemetry and chat, alongside a central middleware handling LA processing and context-aware LLM prompt enrichment. 
This architecture enables the capture of integrated code interaction and chat data, facilitating real-time, context-sensitive AI scaffolding and research into student behaviour. 
We describe the system's design, implementation, and demonstrate its feasibility through system performance benchmarks and two proof-of-concept use cases illustrating its capabilities for logging multi-modal data, analysing help-seeking patterns, and supporting A/B testing of AI configurations. 
JELAI's primary contribution is its technical framework, providing a flexible tool for researchers and educators to develop, deploy, and study LA-informed AI tutoring within the widely used Jupyter ecosystem.

\keywords{AI devices and tools \and Pedagogy and LLMs \and Learning Analytics \and Open-source}
\vspace{-0.4cm}

\end{abstract}



\section{Introduction} 
Generative AI (GenAI) tools like ChatGPT offer potential for on-demand student support, but are generally designed to solve problems rather than support learning pedagogically \cite{giannakos_promise_2024}. 
They typically lack insight into students' learning context (e.g., task, progress, challenges) and can foster reliance on quick answers over deeper understanding \cite{fan_beware_2024}. 
Novices struggle to provide effective context \cite{kazemitabaar_how_2024}, leading to AI responses that may be unhelpful or misaligned with instructional goals \cite{pal_chowdhury_autotutor_2024}. 
Furthermore, research on GenAI in education often relies on self-reports or high-level comparisons, lacking detailed analysis of student-AI interactions needed to understand learning mechanisms and design effective, pedagogically-grounded support \cite{adams_novice_2024,zhang_systematic_2024}.

Integrating Learning Analytics (LA) with GenAI offers a path forward \cite{su_uncovering_2024}. 
Digital environments like Jupyter Notebooks capture rich interaction logs \cite{kitto_creating_2020}. 
Analyzing these logs alongside chat interactions can provide the context needed for adaptive, real-time interventions \cite{valle_torre_sequence_2024,liu_collaboration_2024} and enable research into student cognitive and metacognitive behaviours \cite{fan_beware_2024,ko_trees_2024}. 
While Jupyter is widely used \cite{cardoso_using_2019}, existing extensions typically focus on isolated functions like grading (NBGrader\footnote{\href{https://nbgrader.readthedocs.io/en/stable/}{https://nbgrader.readthedocs.io/en/stable/}}) or provide basic logging/tutoring without deep LA integration \cite{chen_gptutor_2023}. 
Recognizing the need for pedagogically-sound AI in programming, researchers have developed specialized systems like CodeTutor \cite{lyu_evaluating_2024}, CodeAid \cite{kazemitabaar_codeaid_2024}, and CodeHelp \cite{sheese_patterns_2024}, which use techniques like targeted prompting and avoiding direct solutions.
However, these systems often operate separately from the primary coding environment, and their design typically does not focus on the analysis of integrated student workflows within that environment.
To enable such integrated analysis, there is a need for systems that tightly couple fine-grained LA (capturing detailed coding processes and errors) with context-aware conversational AI tutoring \textit{within} the notebook environment.
This integration could bridge the gap between flexible LLMs and more structured learning systems, enabling personalized feedback based on student models and promoting more effective help-seeking \cite{aleven_help_2016}.

\begin{figure}[t!] 
    \centering
    \includegraphics[width=0.9\linewidth]{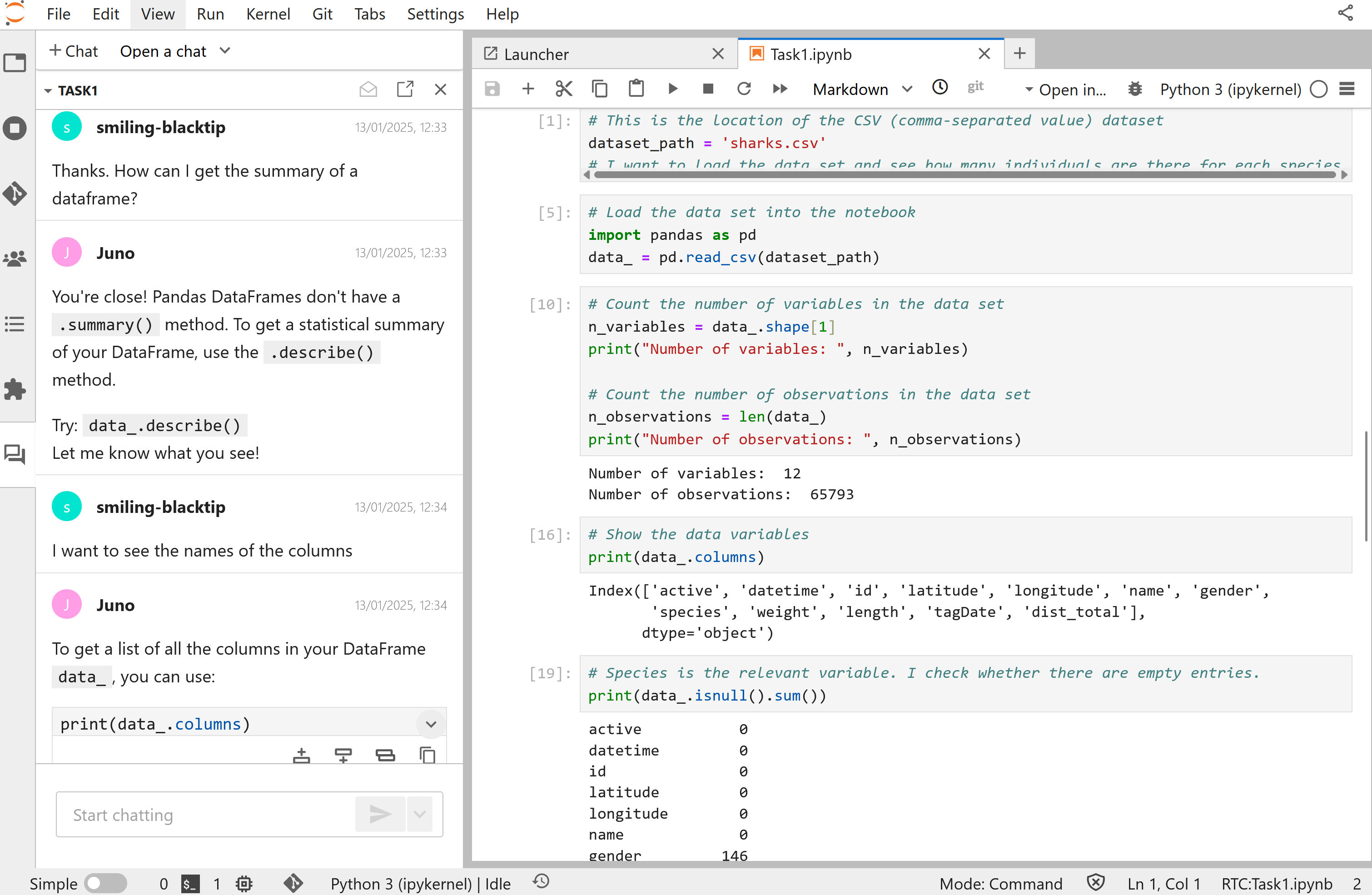} 
    \caption{JELAI Interface: Jupyter Notebook with integrated AI tutor (Juno).}
    \label{fig:screenshot}
\end{figure}

To address this gap, we present JELAI (\textit{Jupyter Environment for Learning Analytics and AI}), an experimental platform integrating LLM-based tutoring with an LA pipeline inside Jupyter notebooks (Figure \ref{fig:screenshot}). 
JELAI is designed as a modular, extensible framework to:
\begin{enumerate}
    \item Capture and process fine-grained student activity (code edits, executions, errors) and chat interactions into meaningful learning traces.
    \item Enrich LLM prompts with real-time context (e.g., recent code, errors, objectives, chat history) for adaptive scaffolding.
    \item Provide a platform for educational research on student-AI interaction patterns and the impact of different AI configurations (e.g., prompting strategies), leveraging the integrated nature of the interaction data.
\end{enumerate}
This paper details JELAI's architecture, implementation, and demonstrates its feasibility through preliminary system performance data and two proof-of-concept use cases. 
The primary contribution is the technical design and implementation of a platform enabling LA-informed AI tutoring within a widely used educational environment.
The open-source repository (BSD-3-Clause license) is available for use and collaboration \footnote{\href{https://github.com/mvallet91/JELAI}{https://github.com/mvallet91/JELAI}}.

\section{System Design and Implementation} 
The design of JELAI was guided by several key requirements necessary for integrating LA with AI tutoring effectively within an educational context:
\begin{itemize}
    \item \textbf{Granular Data Logging:} Capture fine-grained interaction data (code edits, executions, errors, chat) as the foundation for LA and context-aware AI \cite{kitto_creating_2020,su_uncovering_2024}.
    \item \textbf{Real-time Context Enrichment:} Leverage logged data to provide LLMs with immediate context (recent code, errors, conversation history, task goals) to improve relevance and mitigate generic responses \cite{pal_chowdhury_autotutor_2024}.
    \item \textbf{Pedagogical Alignment \& Scaffolding:} Allow instructors to configure AI behaviour (via system prompts, intervention strategies) to align with learning objectives and enable adaptive scaffolding based on student activity \cite{aleven_help_2016,ko_trees_2024}.
    \item \textbf{Modularity and Extensibility:} Design components (logging, LA, LLM interaction) independently to facilitate updates, integration of new techniques, and research experimentation.
    \item \textbf{Scalability and Privacy:} Support multiple concurrent users efficiently while ensuring data isolation and privacy, particularly when using local LLMs.
\end{itemize}
To meet these requirements, JELAI is implemented as a modular, containerized system using open-source technologies (Docker, JupyterHub, JupyterLab, Ollama) and custom Jupyter extensions (\texttt{Jupyter-Chat}, \texttt{JupyterLab-Pioneer}).
JELAI is deployed using \texttt{docker-compose}, simplifying setup and ensuring user isolation for scalability and privacy.

\begin{figure}[h]
    \centering
    \includegraphics[width=0.7\linewidth]{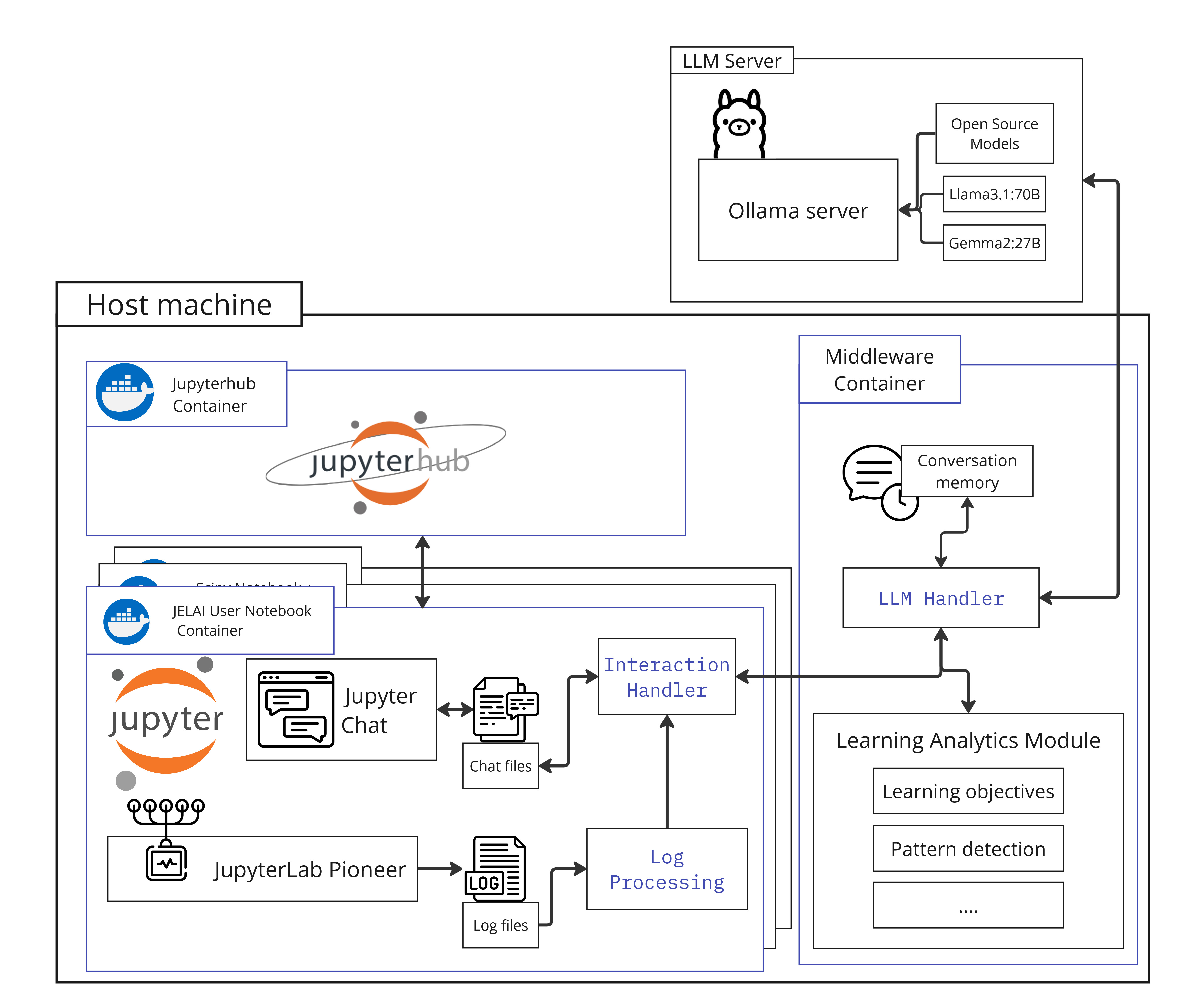}
    \caption{JELAI System Architecture.}
    \label{fig:architecture}
\end{figure}

The architecture (Figure \ref{fig:architecture}) comprises four main components:
\begin{enumerate}
    \item \textbf{User Notebook Container:} Hosts the student's JupyterLab instance. 
    It includes the \texttt{Jupyter-Chat}\footnote{\href{https://github.com/jupyterlab/jupyter-chat}{https://github.com/jupyterlab/jupyter-chat}} extension for the AI tutor interface and the \texttt{JupyterLab-Pioneer}\footnote{\href{https://jupyterlab-pioneer.readthedocs.io/en/latest/}{https://jupyterlab-pioneer.readthedocs.io/en/latest/}} extension for capturing detailed telemetry (keystrokes, cell executions, errors, UI interactions). 
    An \texttt{Interaction Handler} within this container preprocesses student chat messages, enriching them with immediate context (e.g., recent edits and errors, task objective) before forwarding them.
    \item \textbf{Middleware Container:} Acts as the central orchestration hub. 
    It contains the \texttt{Learning Analytics Module}, which processes incoming telemetry logs and chat data, potentially storing them or making them available for real-time analysis. 
    It also houses the \texttt{LLM Handler}, responsible for retrieving relevant conversation history and context from the LA module, applying instructor-defined pedagogical rules or scaffolding logic (e.g., modifying prompts based on recent student errors or help-seeking patterns), constructing the final prompt, and managing communication with the LLM Server.
    \item \textbf{JupyterHub Container:} Manages user authentication, spawns individual User Notebook Containers, and handles proxying, enabling multi-user deployment.
    \item \textbf{LLM Server:} Provides the interface to the chosen language model. JELAI uses Ollama to support various local open-source LLMs, ensuring data privacy and control, but can also be configured to connect to external APIs compatible with the OpenAI standard.
\end{enumerate}

A typical interaction workflow proceeds as follows: (1) A student interacts within their JupyterLab notebook (e.g., writing code, running cells). 
(2) \texttt{JupyterLab-Pioneer} logs these actions and sends them asynchronously to the Middleware's LA Module. 
(3) The student sends a message to the AI tutor via \texttt{Jupyter-Chat}. 
(4) The local \texttt{Interaction Handler} intercepts the message, adds immediate context (e.g., code from the current cell), and sends it to the Middleware. 
(5) The Middleware's \texttt{LLM Handler} retrieves the message, fetches relevant context from the LA Module, applies any configured pedagogical logic or prompt engineering, and sends the final, context-enriched prompt to the LLM Server.
(6) The LLM generates a response, which is sent back through the Middleware to the \texttt{Jupyter-Chat} interface for the student. 
This architecture ensures that AI responses are informed by both immediate actions and broader interaction history, allowing for more adaptive and pedagogically relevant support. 

\section{Proof-of-Concept Demonstration} 
To demonstrate JELAI's feasibility and capabilities for capturing interaction data and enabling comparative studies, we conducted preliminary evaluations focusing on system performance and two use cases. 
These serve as proof-of-concept illustrations rather than conclusive studies, given the small sample sizes inherent in initial system testing.

\subsection{System Performance}
JELAI demonstrated stable performance on an A40 GPU. 
With local LLMs, such as Llama3.1-70b and Gemma2-27b, average response latencies were 5-9s and 2-3s, respectively, acceptable for interactive use. 
Telemetry processing occurred in near real-time. 
The JupyterHub deployment supported 20+ concurrent users on a moderate server (6 CPU cores, 16GB RAM) with low per-user resource consumption (CPU < 0.25 cores, RAM ~400MB steady state), indicating potential for scalability.

\subsection{Use Case Examples}
Two small-scale studies illustrate JELAI's data collection and experimental capabilities.

\textbf{1. Help-Seeking in Python Course (N=18):} JELAI was used over 7 weeks in an introductory Python course. 
It logged student code interactions and chat messages with an LLM tutor (Llama3.1-70b). 
We categorized chat prompts into instrumental vs. executive help-seeking \cite{nelson-le_gall_help-seeking_1981}. 
\textit{Demonstration:} JELAI successfully captured granular, integrated data streams (code activity, chat logs, help-seeking types). 
This capability allows for exploring relationships between coding behaviours, help-seeking patterns (e.g., relative frequency of executive vs. instrumental requests), and external measures like grades, as well as identifying specific interaction sequences (e.g., help-avoidance despite errors, post-completion verification checks) for qualitative analysis.

\textbf{2. Prompt Comparison Pilot (N=19):} This study demonstrated JELAI's use for A/B testing AI configurations. 
Novice programmers worked on 4 data science tasks using JELAI with either a generic "helpful assistant" prompt or a "pedagogical" prompt (Gemma2-27b).
\textit{Demonstration:} JELAI facilitated the comparison by logging interactions for both groups and automated classification of learners. 
The pedagogical prompt increased dialogue length (17.7 vs. 10.7 messages) and engaged in primarily instrumental (27.9\% vs. 59.6\%), rather than executive requests.  Programming logs allowed us to hypothesize: unprompted students ran more code (12.8 vs.\ 8.3 executions) and encountered more errors (7.4 vs.\ 5.3), a behaviour consistent with productive struggle. 

Together, the pilots confirm that JELAI sustains real‑time responsiveness, surfaces actionable help‑seeking signals, and enables real-time behaviour analysis and rapid A/B testing of prompt designs without infrastructure changes.

\section{Discussion and Future Work} 
This paper presented JELAI, a novel, open-source technical platform integrating LA and AI tutoring within Jupyter Notebooks. 
Its modular architecture allows context-aware AI interaction and, crucially, supports research into student behaviour with GenAI by enabling flexible configuration and data capture.
The proof-of-concept demonstration confirmed system stability and its capability to capture rich, multi-modal data for analyzing interaction patterns and comparing pedagogical strategies, although the preliminary studies require validation with larger samples.

Key limitations include the computational cost of powerful local LLMs (though rapidly improving models mitigate this) and the technical expertise currently needed for configuration. 
Future work will focus on:
1) Enhancing AI interaction with multi-step processing for real-time help classification and proactive interventions \cite{chen_unpacking_2025}.
2) Simplifying configuration via templates and higher-level indicators \cite{kerimbayev_intelligent_2025}.
3) Integrating RAG and AI Agents for improved tutoring and tool use.
4) Improving interoperability via standards like xAPI and LTI \cite{kitto_creating_2020}.
5) Enabling student collaboration features.
Research will further investigate how JELAI can foster instrumental help-seeking and whether guided AI dialogue improves learning outcomes.
In summary, JELAI's main contribution is its flexible technical framework, designed to enable pedagogically-informed, LA-driven AI tutoring research and development within a widely used educational environment, with ongoing work focused on enhancing its usability, capabilities, and interoperability.


%
%
\bibliographystyle{splncs04}
\bibliography{references}

\begin{thebibliography}{10}
\providecommand{\url}[1]{\texttt{#1}}
\providecommand{\urlprefix}{URL }
\providecommand{\doi}[1]{https://doi.org/#1}

\bibitem{adams_novice_2024}
Adams, D., Chuah, K.M., Devadason, E., Azzis, M.S.A.: From novice to navigator: {Students}’ academic help-seeking behaviour, readiness, and perceived usefulness of {ChatGPT} in learning. Education and Information Technologies  \textbf{29}(11),  13617--13634 (Aug 2024). \doi{10.1007/s10639-023-12427-8}, \url{https://doi.org/10.1007/s10639-023-12427-8}

\bibitem{aleven_help_2016}
Aleven, V., Roll, I., McLaren, B.M., Koedinger, K.R.: Help {Helps}, {But} {Only} {So} {Much}: {Research} on {Help} {Seeking} with {Intelligent} {Tutoring} {Systems}. International Journal of Artificial Intelligence in Education  \textbf{26}(1),  205--223 (Mar 2016). \doi{10.1007/s40593-015-0089-1}, \url{https://doi.org/10.1007/s40593-015-0089-1}

\bibitem{cardoso_using_2019}
Cardoso, A., Leitão, J., Teixeira, C.: Using the {Jupyter} {Notebook} as a {Tool} to {Support} the {Teaching} and {Learning} {Processes} in {Engineering} {Courses}. In: Auer, M.E., Tsiatsos, T. (eds.) The {Challenges} of the {Digital} {Transformation} in {Education}. pp. 227--236. Springer International Publishing, Cham (2019). \doi{10.1007/978-3-030-11935-5_22}

\bibitem{chen_unpacking_2025}
Chen, A., Xiang, M., Zhou, J., Jia, J., Shang, J., Li, X., Gašević, D., Fan, Y.: Unpacking help-seeking process through multimodal learning analytics: {A} comparative study of {ChatGPT} vs {Human} expert. Computers \& Education  \textbf{226},  105198 (Mar 2025). \doi{10.1016/j.compedu.2024.105198}, \url{https://www.sciencedirect.com/science/article/pii/S0360131524002124}

\bibitem{chen_gptutor_2023}
Chen, E., Huang, R., Chen, H.S., Tseng, Y.H., Li, L.Y.: {GPTutor}: {A} {ChatGPT}-{Powered} {Programming} {Tool} for {Code} {Explanation}. In: Artificial {Intelligence} in {Education}. {Posters} and {Late} {Breaking} {Results}, {Workshops} and {Tutorials}, {Industry} and {Innovation} {Tracks}, {Practitioners}, {Doctoral} {Consortium} and {Blue} {Sky}. pp. 321--327. Springer, Cham (2023). \doi{10.1007/978-3-031-36336-8_50}, \url{https://link.springer.com/chapter/10.1007/978-3-031-36336-8_50}, iSSN: 1865-0937

\bibitem{fan_beware_2024}
Fan, Y., Tang, L., Le, H., Shen, K., Tan, S., Zhao, Y., Shen, Y., Li, X., Gašević, D.: Beware of metacognitive laziness: {Effects} of generative artificial intelligence on learning motivation, processes, and performance. British Journal of Educational Technology  \textbf{n/a}(n/a) (Dec 2024). \doi{10.1111/bjet.13544}, \url{https://onlinelibrary.wiley.com/doi/abs/10.1111/bjet.13544}

\bibitem{giannakos_promise_2024}
Giannakos, M., Azevedo, R., Brusilovsky, P., Cukurova, M., Dimitriadis, Y., Hernandez-Leo, D., Järvelä, S., Mavrikis, M., Rienties, B.: The promise and challenges of generative {AI} in education. Behaviour \& Information Technology  \textbf{0}(0),  1--27 (2024). \doi{10.1080/0144929X.2024.2394886}, \url{https://doi.org/10.1080/0144929X.2024.2394886}

\bibitem{kazemitabaar_how_2024}
Kazemitabaar, M., Hou, X., Henley, A., Ericson, B.J., Weintrop, D., Grossman, T.: How {Novices} {Use} {LLM}-based {Code} {Generators} to {Solve} {CS1} {Coding} {Tasks} in a {Self}-{Paced} {Learning} {Environment}. In: Proceedings of the 23rd {Koli} {Calling} {International} {Conference} on {Computing} {Education} {Research}. pp. 1--12. Koli {Calling} '23, Association for Computing Machinery, New York, NY, USA (Feb 2024). \doi{10.1145/3631802.3631806}, \url{https://dl.acm.org/doi/10.1145/3631802.3631806}

\bibitem{kazemitabaar_codeaid_2024}
Kazemitabaar, M., Ye, R., Wang, X., Henley, A.Z., Denny, P., Craig, M., Grossman, T.: {CodeAid}: {Evaluating} a {Classroom} {Deployment} of an {LLM}-based {Programming} {Assistant} that {Balances} {Student} and {Educator} {Needs}. In: Proceedings of the 2024 {CHI} {Conference} on {Human} {Factors} in {Computing} {Systems}. pp. 1--20. {CHI} '24, Association for Computing Machinery, New York, NY, USA (May 2024). \doi{10.1145/3613904.3642773}, \url{https://dl.acm.org/doi/10.1145/3613904.3642773}

\bibitem{kerimbayev_intelligent_2025}
Kerimbayev, N., Adamova, K., Shadiev, R., Altinay, Z.: Intelligent educational technologies in individual learning: a systematic literature review. Smart Learning Environments  \textbf{12}(1), ~1 (Jan 2025). \doi{10.1186/s40561-024-00360-3}, \url{https://doi.org/10.1186/s40561-024-00360-3}

\bibitem{kitto_creating_2020}
Kitto, K., Whitmer, J., Silvers, A., Webb, M.: Creating data for learning analytics ecosystems [position paper]. Report, Society for Learning Analytics Research (Sep 2020), \url{https://opus.lib.uts.edu.au/handle/10453/152209}, accepted: 2021-12-08T00:01:48Z

\bibitem{ko_trees_2024}
Ko, S.H., Stephens-Martinez, K.: The {Trees} in the {Forest}: {Characterizing} {Computing} {Students}' {Individual} {Help}-{Seeking} {Approaches}. In: Proceedings of the 2024 {ACM} {Conference} on {International} {Computing} {Education} {Research} - {Volume} 1. {ICER} '24, vol.~1, pp. 343--358. Association for Computing Machinery, New York, NY, USA (Aug 2024). \doi{10.1145/3632620.3671099}, \url{https://dl.acm.org/doi/10.1145/3632620.3671099}

\bibitem{liu_collaboration_2024}
Liu, J., Li, S., Dong, Q.: Collaboration with {Generative} {Artificial} {Intelligence}: {An} {Exploratory} {Study} {Based} on {Learning} {Analytics}. Journal of Educational Computing Research  \textbf{62}(5),  1234--1266 (Sep 2024). \doi{10.1177/07356331241242441}, \url{https://doi.org/10.1177/07356331241242441}, publisher: SAGE Publications Inc

\bibitem{lyu_evaluating_2024}
Lyu, W., Wang, Y., Chung, T.R., Sun, Y., Zhang, Y.: Evaluating the {Effectiveness} of {LLMs} in {Introductory} {Computer} {Science} {Education}: {A} {Semester}-{Long} {Field} {Study}. In: Proceedings of the {Eleventh} {ACM} {Conference} on {Learning} @ {Scale}. pp. 63--74. L@{S} '24, Association for Computing Machinery, New York, NY, USA (Jul 2024). \doi{10.1145/3657604.3662036}, \url{https://dl.acm.org/doi/10.1145/3657604.3662036}

\bibitem{nelson-le_gall_help-seeking_1981}
Nelson-Le~Gall, S.: Help-seeking: {An} understudied problem-solving skill in children. Developmental Review  \textbf{1}(3),  224--246 (Sep 1981). \doi{10.1016/0273-2297(81)90019-8}, \url{https://www.sciencedirect.com/science/article/pii/0273229781900198}

\bibitem{pal_chowdhury_autotutor_2024}
Pal~Chowdhury, S., Zouhar, V., Sachan, M.: {AutoTutor} meets {Large} {Language} {Models}: {A} {Language} {Model} {Tutor} with {Rich} {Pedagogy} and {Guardrails}. In: Proceedings of the {Eleventh} {ACM} {Conference} on {Learning} @ {Scale}. pp. 5--15. L@{S} '24, Association for Computing Machinery, New York, NY, USA (Jul 2024). \doi{10.1145/3657604.3662041}, \url{https://dl.acm.org/doi/10.1145/3657604.3662041}

\bibitem{sheese_patterns_2024}
Sheese, B., Liffiton, M., Savelka, J., Denny, P.: Patterns of {Student} {Help}-{Seeking} {When} {Using} a {Large} {Language} {Model}-{Powered} {Programming} {Assistant}. In: Proceedings of the 26th {Australasian} {Computing} {Education} {Conference}. pp. 49--57. {ACE} '24, Association for Computing Machinery, New York, NY, USA (Jan 2024). \doi{10.1145/3636243.3636249}, \url{https://dl.acm.org/doi/10.1145/3636243.3636249}

\bibitem{su_uncovering_2024}
Su, H., Tong, Y., Zhang, X., Fan, Y.: Uncovering {Students}’ {Processing} {Tactics} {Towards} {ChatGPT}’s {Feedback} in {EFL} {Education} {Using} {Learning} {Analytics}. In: Ma, W.W.K., Li, C., Fan, C.W., U, L.H., Lu, A. (eds.) Blended {Learning}. {Intelligent} {Computing} in {Education}. pp. 238--250. Springer Nature, Singapore (2024). \doi{10.1007/978-981-97-4442-8_18}

\bibitem{valle_torre_sequence_2024}
Valle~Torre, M., Oertel, C., Specht, M.: The {Sequence} {Matters} in {Learning} - {A} {Systematic} {Literature} {Review}. In: Proceedings of the 14th {Learning} {Analytics} and {Knowledge} {Conference}. pp. 263--272. {LAK} '24, Association for Computing Machinery, New York, NY, USA (Mar 2024). \doi{10.1145/3636555.3636880}, \url{https://doi.org/10.1145/3636555.3636880}

\bibitem{zhang_systematic_2024}
Zhang, X., Zhang, P., Shen, Y., Liu, M., Wang, Q., Gašević, D., Fan, Y.: A {Systematic} {Literature} {Review} of {Empirical} {Research} on {Applying} {Generative} {Artificial} {Intelligence} in {Education}. Frontiers of Digital Education  \textbf{1}(3),  223--245 (Sep 2024). \doi{10.1007/s44366-024-0028-5}, \url{https://doi.org/10.1007/s44366-024-0028-5}

\end{thebibliography}
\end{document}